\documentclass[aps,prl,twocolumn,showpacs,amsmath,amssymb,superscriptaddress]{revtex4-1}

\usepackage{amsfonts}
\usepackage{amsmath}
\usepackage{graphicx}
\usepackage{bm}
\usepackage{amssymb}
\usepackage{lipsum}
\usepackage{times}
\usepackage{dcolumn}
\usepackage{cases}
\usepackage{braket}
\usepackage{booktabs}
\usepackage{setspace} 
\usepackage{color}

\usepackage{array}
\usepackage{float}
\usepackage{blindtext}

\makeatletter
    \renewcommand\@make@capt@title[2]{%
     \@ifx@empty\float@link{\@firstofone}{\expandafter\href\expandafter{\float@link}}%
      {\textbf{#1}}\@caption@fignum@sep#2\quad}%
    \makeatother
   
\makeatletter 
\renewcommand{\fnum@figure}{\textbf{Figure~\thefigure}}
\makeatother

\begin{document}

\title{Dzyaloshinskii-Moriya interaction and Hall effects in the skyrmion phase of Mn$_{1-x}$Fe$_x$Ge}
\author{J. Gayles}
\affiliation{Institut f\"ur Physik, Johannes Gutenberg Universit\"at Mainz, D-55099 Mainz, Germany}
\affiliation{Department of Physics \& Astronomy, Texas A\&M University,
College Station, Texas 77843-4242, USA}
\author{F. Freimuth}
\affiliation{  Peter Gr{\"u}nberg Institut \& Institute for Advanced Simulation, Forschungszentrum J{\"u}lich and JARA, 52425 J{\"u}lich, Germany}
\author{T. Schena}
\affiliation{  Peter Gr{\"u}nberg Institut \& Institute for Advanced Simulation, Forschungszentrum J{\"u}lich and JARA, 52425 J{\"u}lich, Germany}
\author{G. Lani}
\affiliation{  Peter Gr{\"u}nberg Institut \& Institute for Advanced Simulation, Forschungszentrum J{\"u}lich and JARA, 52425 J{\"u}lich, Germany}
\author{P. Mavropoulos}
\affiliation{  Peter Gr{\"u}nberg Institut \& Institute for Advanced Simulation, Forschungszentrum J{\"u}lich and JARA, 52425 J{\"u}lich, Germany}
\author{R. Duine}
\affiliation{Institute for Theoretical Physics, Utrecht University, Leuvenlaan 4, 3584 CE Utrecht, The Netherlands}
\author{S. Bl{\"u}gel}
\affiliation{  Peter Gr{\"u}nberg Institut \& Institute for Advanced Simulation, Forschungszentrum J{\"u}lich and JARA, 52425 J{\"u}lich, Germany}
\author{J. Sinova}
\affiliation{Institut f\"ur Physik, Johannes Gutenberg Universit\"at Mainz, D-55099 Mainz, Germany}
\affiliation{Department of Physics \& Astronomy, Texas A\&M University, College Station, Texas 77843-4242, USA}
\affiliation{Institute of Physics ASCR, v.v.i., Cukrovarnicka 10, 162 53 Praha 6 Czech Republic}
\author{Y. Mokrousov}
\affiliation{  Peter Gr{\"u}nberg Institut \& Institute for Advanced Simulation, Forschungszentrum J{\"u}lich and JARA, 52425 J{\"u}lich, Germany}

\begin{abstract}
We carry out density functional theory calculations which demonstrate that the electron dynamics in the skyrmion phase of Fe-rich Mn$_{1-x}$Fe$_x$Ge alloys is governed by Berry phase physics.  We observe that the magnitude of the Dzyaloshinskii-Moriya interaction, directly related to the mixed space-momentum Berry phases, changes sign and magnitude with concentration $x$ in direct correlation with the data of Shibata {\it et al.}, Nature Nanotech. {\bf 8}, 723 (2013). 
The computed anomalous and topological Hall effects in FeGe are also in good agreement with available experiments.
We further develop a simple tight-binding model able to explain these findings. Finally, we show that
the adiabatic Berry phase picture is violated in the Mn-rich limit of the alloys.
\end{abstract}

\maketitle

Recently, there has been strong interest in skyrmionic systems for applications in spintronic devices. Skyrmions in magnetic systems are 
whirls of magnetization that have a non-zero topological charge, also known as the winding number.
These topologically protected structures are particularly promising in magnetic memory devices \cite{Wiesen}, where memory bits can be packed denser  and are more robust due to their topological nature. 
In addition, it has been experimentally shown that current densities used to manipulate these particle-like magnetic whirls
are five orders of magnitude lower than in magnetic switching devices based on spin-transfer torque~\cite{Science2010,Fert}. 

Chiral skyrmions were first seen to exist 
in the so-called B20 compounds, of which the most prominent representatives are 
MnSi, FeCoSi, FeGe and MnGe based alloys~\cite{Muh,tok2,Wilh}. What makes the B20
materials so special is the real space inversion asymmetry, itinerant magnetism and often relatively small spin-orbit interaction (SOI). The electronic
and magnetic properties of these alloys are very sensitive to various parameters, such as pressure,
temperature and alloy composition. 
The phase diagram of many B20 compounds with respect to temperature and magnetic field consists of several phases. Most importantly, 
it often exhibits the $A$-phase characterized by formation of a chiral skyrmion lattice below a critical 
temperature in a finite external field~\cite{Muh,Science2007}.   
Recently, it was shown experimentally that in Mn$_{1-x}$Fe$_x$Ge alloys the skyrmions in the $A$-phase 
drastically change their size and chirality as a function of chemical composition ~\cite{tokura1,Russian1}. 
 
The fundamental interaction behind the formation of chiral skyrmions in B20 compounds is the antisymmetric Dzyaloshinskii-Moriya exchange interaction (DMI)~\cite{M1,M2,M3,M11,M5,M6}. The DMI arises in crystals with broken inversion symmetry and it favors a certain chirality of the magnetization $-$ the condition,
necessary for formation of chiral magnetic structures such as skyrmions or spin-spirals of unique rotational sense. For slowly varying  magnetic textures the contribution to the total energy of the system due to the DMI reads  $E_{DM}=\sum_{i}\,{\bf D}_i ({\bf \hat{m}})\cdot \left( {\bf \hat{m}} \times \partial_i {\bf \hat{m}} \right)$, where $i$ stands for cartesian coordinates, ${\bf D}_i$ is the $i$'th Dzyaloshinskii-Moriya vector, and $\bf \hat{m}$ is the unit vector of the space-dependent magnetization. The DMI has been known since the 1950s from symmetry grounds, 
yet the physics which dictate its properties in transition-metal compounds remain largely unexplored. 
Recently, it was shown that in geometric terms the DMI is intrinsically related to the so-called
mixed part of the Berry curvature (BC) tensor which couples the real- and reciprocal space evolution of the electronic states  in chiral skyrmion lattices with weak SOI~\cite{Berry}. 
As was unambiguously demonstrated for Mn$_x$Fe$_{1-x}$Si alloys~\cite{Franktop,mnsithe}, the transport facets of the purely reciprocal- and real-space BC are the anomalous 
Hall (AHE) and the topological Hall (THE) effects, respectively. Of the two Hall effects,  the THE in particular plays a crucial role in detection of skyrmions by electrical means~\cite{mnsithe}.

In this Letter, using first principles techniques and connecting to recent experiments, we show that 
the adiabatic Berry picture governs the electron dynamics in the Fe-rich Mn$_x$Fe$_{1-x}$Ge alloys.
This not only applies to the real-space and reciprocal-space Berry phases as seen
from the agreement between the calculated THE and AHE and experiments on FeGe, 
but also to the effects of the mixed Berry phases as manifested by the dependence of the Dzyaloshinskii-Moriya interaction on the Fe concentration. 
Namely, the change of sign of the DMI at the critical concentration of $x=0.8$ in Mn$_x$Fe$_{1-x}$Ge  is
in excellent agreement to observations reported in Ref.~\cite{tokura1,Russian1}. To further understand our findings,
guided by {\it ab-initio} insight, we develop a minimal tight-binding model of the DMI, which 
accounts for its peculiar sign change. We further show that the limits of the adiabatic Berry
phase paradigm are not met at the Mn-rich side of Mn$_x$Fe$_{1-x}$Ge alloys. 
Our findings should help 
the material design of systems which exhibit skyrmionic states.

We have carried out density functional theory (DFT) calculations of bulk Mn$_{1-x}$Fe$_x$Ge alloys using the full-potential linearized augmented plane wave method as implemented in the J\"ulich DFT code \texttt{FLEUR}~\cite{fleur}, and the  Perdew-Burke-Ernzerhof (PBE)~\cite{PBE} parametrization of the
exchange-correlation potential. To treat the effect of disorder we employed the virtual crystal approximation (VCA)~\cite{VCA}. Starting from the experimental lattice constants of pure MnGe~\cite{PhysRevB.85.205205}
and FeGe~\cite{Wilhelm} we  used Vegard's lattice constants for $0<x<1$. The collinear ferromagnetic calculations yield a magnetic moment of $2.2\,\mu_B$ and $1.2\,\mu_B$ in MnGe and FeGe respectively, which compare well with the corresponding experimental values of $2.3\,  \mu_B$ and $1.0\,\mu_B$~\cite{Wilhelm}.
More details on computation of the electronic structure, transport properties and setup of the minimal
tight-binding model are given in Supplementary Information. 

When computing the DMI we neglect the anisotropy of the DMI vector with respect to $\hat{\bf m}$, which we have found to be very small in the studied alloys.
In this case the impact of the DMI can be accounted for by a single constant $D$, which characterizes an
energy difference between the flat (non-conical) spin-spiral states of opposite rotational sense. Changing the sign of $D$
would result in a change of the rotational sense of the energetically preferred spin-spiral solution. 
To compute the value of $D$, we used two methods, which gave very similar results for the considered alloys. The first one
is based on the expression for the DMI obtained from the Berry phase theory in the weak SOI limit (Eq. 11 in Ref.~\onlinecite{Berry}). 
The second one is based on evaluating the linear slope of the dispersion energy of the long wavelength flat spin-spiral solutions when
including the SOI within first order perturbation theory (see Supplementary Information)~\cite{Pert}. 
The two methods coincide in the limit of weak SOI strength for cubic crystals. 
In this work, we present the values obtained with the second method, since it
allows for a transparent decomposition of the DMI into contributions coming from different atomic species.

\begin{figure}
    \centering
    \includegraphics[height=6.5cm]{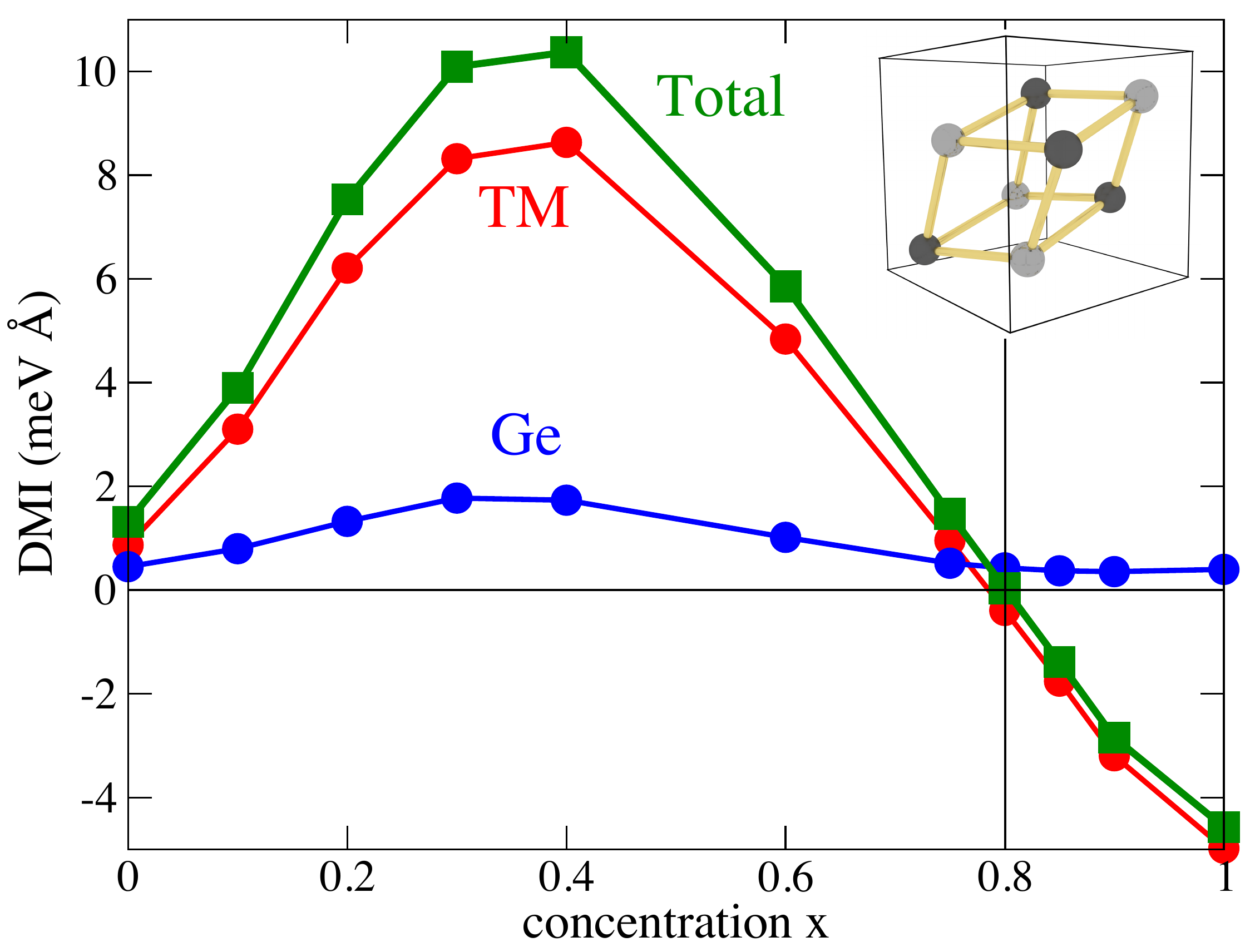}%
    \caption{(Color online) Strength of the DMI as a function of concentration $x$ in Mn$_{1-x}$Fe$_{x}$Ge alloys. The total value of the DMI (filled squares) is decomposed into the contributions coming from the transition-metal (red dots) and Ge (blue dots). The inset depicts the crystal structure of the studied  B20 compound, with light grey and dark grey spheres representing the transition-metal (TM) and Ge atoms, respectively. }
    \label{DMI}
  \end{figure} 

The results of our calculations of the DMI strength $D$ in Mn$_{1-x}$Fe$_{x}$Ge alloys are presented in Fig.~\ref{DMI} as a function of concentration $x$.  We first focus on the Fe-rich  side ($x\rightarrow 1$). Our most remarkable finding is 
the change of the sign of $D$ at the critical concentration $x_c=0.8$, which results in the change of magnetic helicity of skyrmions in excellent agreement with the recent experimental observations in the $A$-phase
of Mn$_{1-x}$Fe$_{x}$Ge alloys~\cite{tokura1}. At $x_c$ the DMI strength $D$ vanishes, which theoretically
should result in an infinite pitch $\lambda_{\rm sk}$ of the skyrmions at this concentration, since 
$\lambda_{\rm sk}\sim J/D$, with $J$ being the Heisenberg exchange in the system~\cite{tokura1,Russian1}. In addition  experiments observe a fall-off law $\lambda_{\rm sk}\sim \left|x_c-x\right |^{-1}$ in the vicinity 
of $x_c$, predicted by our calculations as a direct consequence of the linear behavior of the DMI strength at the critical concentration 
$D\sim(x_c-x)$. The sign of $D$ to the left (positive) and to the right
(negative) of $x_c$, which determines the sense of magnetic helicity, is also in agreement to experiments, given
that the structural chirality of the B20 lattice of our alloys is kept constant as a function of $x$, and
is the same as for MnSi~\cite{PhysRevB.88.024408}.

Within our approximation of disorder, the B20 lattice for $0 < x < 1$ consists of two atomic species: Ge atoms, and effective transition-metal (TM) atoms, whose atomic properties are a mixture of those of Fe and Mn atoms~\cite{VCA}. Our method allows us to decompose the DMI into contributions coming from these two different atomic
species. As seen in Fig.~\ref{DMI}, where this decomposition is presented, the overall trend of the DMI 
as a function of $x$ is almost solely determined by the contribution from the  TM. Since a contribution to the DMI from a given atom is directly proportional to the SOI strength on it, we conclude that it is the SOI
coming from the TM which is responsible for the DMI in this family of alloys.

When decreasing the concentration away from $x_c$ we first observe a rapid increase of $D$, which reaches as much as~10\,meV\,\AA~at $x=0.4$. This is in agreement with experiments as well, which predict a rapid decrease of 
$\lambda_{\rm sk}$ with increasing $|x-x_c|$. 
This is confirmed by our calculations for which the Heisenberg exchange interactions do not change significantly when going from pure FeGe to MnGe, and thus the relation $\lambda_{\rm sk}\sim 1/D$ should be satisfied. 
However, upon further decreasing 
$x$, the DMI strength decreases, constituting a small value of~1.2\,meV\,\AA~for MnGe. 
Thus, close to 
pure MnGe we are unable to explain the experimental finding of monotonously decreasing $\lambda_{\rm sk}$ down to zero with decreasing $x$, resulting in an observation of ultra-small size of skyrmions in MnGe on the order of 3\,nm~\cite{tok3}.
Since we believe that our {\it ab-initio} description of the electronic structure of MnGe is reliable, 
we attribute this discrepancy for  Mn-rich Mn$_{1-x}$Fe$_x$Ge to the breakdown of the assumption of slowly varying magnetization, used to evaluate the $D$,~{\it i.e.},  the breakdown of adiabatic approximation. Another possible explanation for this discrepancy could be that the real spin structure in MnGe is more complex than a simple skyrmion lattice. The very small value of $\lambda_{\rm sk}$ makes current experimental measurement challenging and leaves ambiguity in the structure of the spin lattice in MnGe~\cite{tokura1,Sky2}.

\begin{figure}
    \centering
    \includegraphics[height=6.0cm]{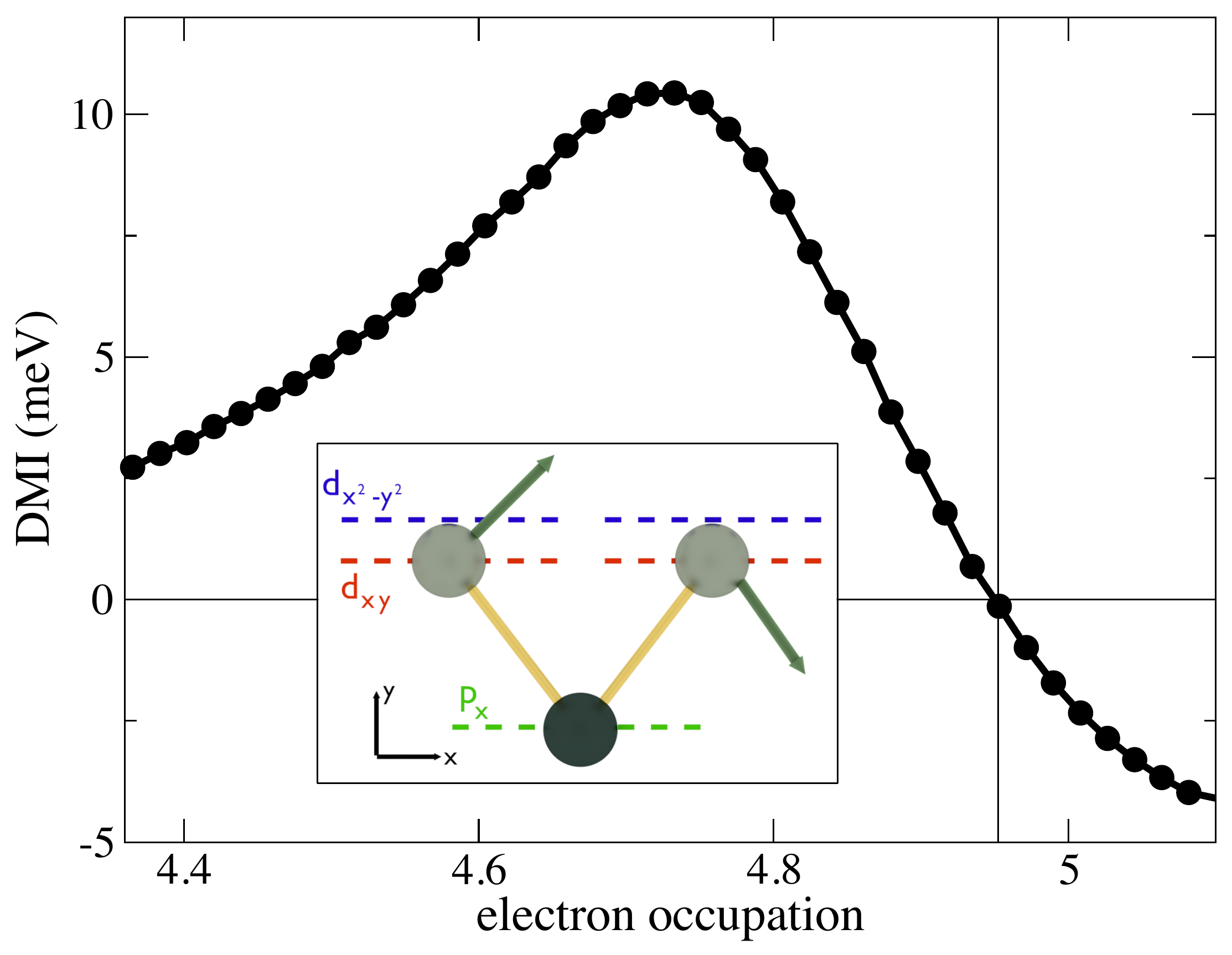}%
\caption{(Color online) Strength of the DMI as a function of electron occupation computed within simple 
tight-binding model of a finite trimer. The inset shows the structure of the trimer designed to mimic the breaking of 
local inversion symmetry of the bond between the transition-metal (light grey) and Ge (dark grey), which
gives rise to the DMI. The  direction of the left ($\mathbf{S}_1$) and right ($\mathbf{S}_2$) spin lies in the $xy$ plane, and the DMI vector is pointing out-of-plane.}
 \label{tDMI}
 \end{figure}

To understand the origin of the sign change in the DMI we develop a  minimal tight-binding model for a finite trimer system (inset in Fig.~\ref{tDMI}), positioned in the $xy$-plane. Within our model, the trimer of atoms mimics the bond between the two TM and one Ge atom in B20 structure~(see inset in Fig.~\ref{DMI}). 
This model is derived in a similar way as our previous model for 3$d$-5$d$ transition metal chains~\cite{timo}, and
it captures the essential physics of the DMI in our Mn$_{1-x}$Fe$_x$Ge alloys. 
Based on the DFT results, in our model we neglect the SOC on the Ge atom, while the effects of non-collinearity
and SOC on TMs lead to a finite DMI strength, $D=|\mathbf{D}|$, via contribution to the energy of the type  $E_{DM}=\mathbf{D}\cdot(\mathbf{S}_1\times\mathbf{S}_2$),
with $\mathbf{S}_1$ and $\mathbf{S}_2$ as spin moments of two TM atoms. The Ge atom is represented with
one $p_x$ orbital per spin (spin-degenerate), while the TM is represented with $d_{xy}$ and $d_{x^2-y^2}$ orbitals per spin (exchange split), and
only Ge $p_x$ and TM $d_{xy}, d_{x^2-y^2}$ orbitals are allowed to have non-zero inter-atomic hopping.
Within this model a finite DMI is estimated from the difference in energy between two configurations of $\mathbf{S}_1$ and $\mathbf{S}_2$: $\mathbf{S}_1$ at an angle of $+\frac{\pi}{4}$ ($-\frac{\pi}{4}$) and $\mathbf{S}_2$ at an angle of $-\frac{\pi}{4}$ ($+\frac{\pi}{4}$) from the $x$-axis, with both spins lying in the $xy$-plane. 
In this setup the vector $\mathbf{D}$ lies out-of-plane.

We mimic the change of concentration $x$ in Mn$_{1-x}$Fe$_x$Ge by changing the electronic occupation of the orbitals,
tuning the change in the spin moment and relative positions of the $p_{x}$, $d_{xy}$ and $d_{x^2-y^2}$ orbitals in 
accordance with first principles calculations of the electronic structure of the  alloys (see Supplementary material for more details).  The results of our model calculations for the DMI strength as a function of $x$, presented in Fig.~\ref{tDMI}, are very similar to those
obtained from first principles,~Fig.~\ref{DMI}. The minimal number of ingredients entering our model help us
pin down the main mechanism behind the peculiar behavior of the DMI in Mn$_{1-x}$Fe$_x$Ge alloys. Namely, it is the dynamics
of the $d_{x^2-y^2}$-like states $-$ which move from above the Fermi energy, become occupied and enter the
region of $d_{xy}$-states of opposite spin with increasing the concentration $x$ $-$ that is responsible for 
the change of sign and local peak in the DMI strength in the vicinity of $x_c$.

\begin{figure}[t!]
\includegraphics[width=1.0\linewidth]{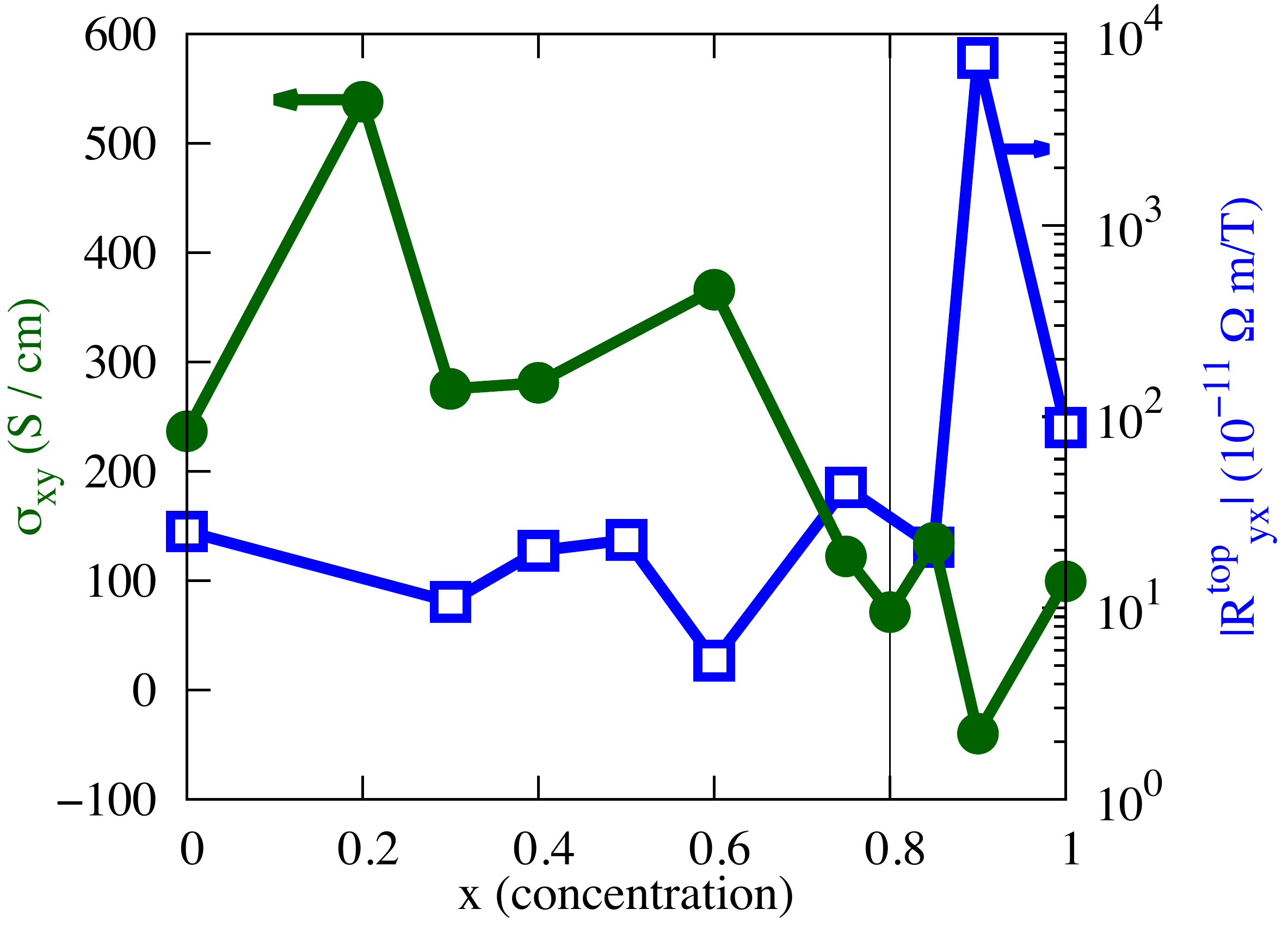}
\caption{(Color online) Computed anomalous Hall conductivity $\sigma_{xy}$ and the absolute value of topological Hall constant $R^{\rm top}_{yx}$ as a function of concentration $x$ in Mn$_{1-x}$Fe$_x$Ge alloys. Thin line
marks the critical concentration $x_c=0.8$.}
\label{top}
\end{figure}

In the language of Berry phases the DMI is directly related to the off-diagonal component of the BC
 tensor $\Omega$ which mixes real ($\mathbf{R}$) and reciprocal ($\mathbf{k}$) spaces~\cite{Berry}. The fact that our calculations for the DMI agree with experiments on Fe-rich Mn$_{1-x}$Fe$_x$Ge suggests the validity of the Berry phase physics in these systems, but also poses a question as to whether such an agreement extends also to the other effects which hinge on the
diagonal components of the BC tensor, namely, the real-space and reciprocal-space BCs, $\Omega_{\mathbf{RR}}$ and $\Omega_{\mathbf{kk}}$, respectively. Important transport manifestations
of the latter two BCs are the anomalous and topological Hall effects, currently studied intensively in skyrmionic
systems~\cite{tok3,mngeco,FeGetop,mngeco,Huangprl,Franktop,mnsithe,mnsithe2,mnsiahe}.
The dynamics of an electron in a given band which travels through a skyrmionic system is
completely determined
by the full BC tensor~\cite{Berry}. The exact expression for,~{\it e.g.}, the $\mathbf{k}$-part of the BC tensor reads 
$\Omega_{\mathbf{kk},ij}
=-2{\rm Im}\Braket{\frac{\partial u}{\partial\mathbf{k}_i}|
\frac{\partial u}{\partial\mathbf{k}_j}}$, where $i$ and $j$ mark Cartesian components. The lattice-periodic part of an electron in the considered band,  $u=u(\mathbf{k},\mathbf{R})$, is computed for a ferromagnetic crystal with the magnetization direction $\hat{\mathbf{m}}(\mathbf{R})$ determined by the position $\mathbf{R}$ within the skyrmion. The other two components of the BC tensor, 
$\Omega_{\mathbf{Rk}}$ and $\Omega_{\mathbf{RR}}$, are computed analogously.

We first consider the reciprocal space and evaluate the $\mathbf{k}$-resolved and summed over all occupied states BC $\Omega_{\mathbf{kk}}$ for [001]-direction of the magnetization in our ferromagnetic Mn$_{1-x}$Fe$_x$Ge crystal for all $x$. Our calculations show that the anisotropy of $\Omega_{\mathbf{kk}}$ with respect to the direction of the magnetization is rather small. The manifestation of $\Omega_{\mathbf{kk}}$ is the intrinsic contribution to the AHE~\cite{Jairoahe}, with the anomalous Hall conductivity (AHC) $\sigma_{xy}$ given by the Brillouin zone integral of the non-vanishing $\mathbf{k}$-space BC component $\Omega_{\mathbf{kk},xy}$. The dependence of the computed AHC on the concentration $x$ in Mn$_{1-x}$Fe$_x$Ge alloys, presented in Fig.~\ref{top}, is  ragged, which is typical for transition-metal ferromagnets upon changing the parameters of the electronic structure. Our values can be directly compared to experimental measurements of the AHC in the ferromagnetic phase of MnGe and FeGe, which constitute 150 and 38\,S/cm, respectively~\cite{tok3,FeGetop}.
Clearly, there is a good qualitative agreement in magnitude, sign and trend between our calculations and
experiments, while the remaining differences can be attributed to,~e.g., extrinsic contributions to the AHE~\cite{jurg,PhysRevB.90.220403}.

The major contribution to the real-space BC can be estimated already without
taking SOI into account owing to the small spin-orbit strength of the studied alloys. In this case $\Omega_{\mathbf{RR}}$ can be computed from the knowledge of the magnetization distribution in the skyrmion as 
$\Omega_{\mathbf{RR},ij}=\pm\frac{1}{2}\hat{\mathbf{m}}\cdot(\partial_{\mathbf{R}_i}\hat{\mathbf{m}}\times\partial_{\mathbf{R}_j}\hat{\mathbf{m}})$, with ``$+$" and ``$-$" for spin-up and spin-down electrons. The effect of the real-space BC is that of the spin-dependent
magnetic field which exerts the Lorentz force, opposite for electrons of opposite spin. The averaged over the skyrmion
magnitude of $\Omega_{\mathbf{RR}}$ is known also as the emergent field, $B_e$, and the resulting Hall effect is called the THE. The 
topological Hall resistivity, $\rho^{\rm top}_{yx}$, can be thus computed from
the spin-resolved diagonal and off-diagonal Hall components of the conductivity tensor $\sigma$ as
\begin{equation}\label{8}
\rho^{\rm top}_{yx}=\frac{\sigma_{xy}^{\rm OHE,\uparrow}-\sigma_{xy}^{\rm OHE,\downarrow}}{(\sigma_{xx}^{\uparrow}+\sigma_{xx}^{\downarrow})^2}\,,
\end{equation}
assuming that the modulation of the magnetization occurs within the $xy$-plane
and the emergent field is pointing along the $z$-axis. In order to access the
conductivities from {\it ab-initio} electronic structure without SOI, we assume the Boltzmann approach within the constant relaxation time approximation~\cite{Berry}. Within
this approximation $\rho^{\rm top}_{yx}$ decomposes into the product of
the emergent field, and the so-called {\it topological Hall constant} $R^\mathrm{top}_{yx}$, $\rho^{\rm top}_{yx}=R^\mathrm{top}_{yx}\,B_e$. The topological 
Hall constant can be determined solely from the electronic structure of a material
without the need for any parameters which characterize the scattering off disorder.  

The absolute value of $R^\mathrm{top}_{yx}$ as a function of $x$ in
Mn$_{1-x}$Fe$_x$Ge is shown in Fig.~\ref{top}. One of the most striking features in this dependence is the change in the magnitude of $R^\mathrm{top}_{yx}$ by
orders of magnitude as $x$ is varied. Such a behavior is pronounced especially
in the vicinity of the critical concentration $x=0.8$, where also the AHE undergoes a change in sign. 
We note that although the
variation of the THE, AHE and the DMI with $x$ is driven by the very same redistribution of the electronic states around the Fermi energy, there is in general  
little correlation between the concentration dependence of the three fundamental phenomena.

For pure alloys, the sign of the THE which we predict agrees with the experimental values. In the case of FeGe the value of $R^\mathrm{top}_{yx}$ constitutes 88$\times10^{-11}\, \Omega$m/T  and compares remarkably well with the experimental value of 72$\times10^{-11}\,\Omega$m/T, computed from the  experimental values for $\rho^{\rm top}_{yx}$ and $B_e$~\cite{FeGetop}. In MnGe we obtain a value of 25$\times10^{-11}\,\Omega$m/T for $R^\mathrm{top}_{yx}$, which is two orders of magnitude larger than the experimental value of 0.4$\times10^{-11}\,\Omega$m/T \,\cite{tok3}.
The overestimation of the topological Hall constant in this case, in analogy to the DMI in this limit, can be attributed to the breakdown in the adiabatic approximation, essential in the Berry phase viewpoint, owing to the inability 
of a conduction spin to follow the rapidly changing magnetization of the
skyrmion lattice. The physics of the electron dynamics and Hall effects in 
this regime, and its proper description with first principles methods, present 
an important direction to tackle, especially in the light of recent intensive
interest in nano-scale non-trivial spin textures arising at surfaces and interfaces~\cite{Dupe2014,nphys2045,HeinzeYM}.

We acknowledge  the financial support from the NSF grant no. DMR-1105512,   the Alexander Von Humboldt
Foundation, the HGF Programme VH-NG 513 and DFG
SPP 1568. RD is supported by the Stichting voor Fundamenteel
Onderzoek der Materie (FOM) and is part of the D-ITP consortium, a program of the Netherlands Organisation for
Scientific Research (NWO) that is funded by the Dutch
Ministry of Education, Culture and Science (OCW). 
 We also gratefully  acknowledge J\"ulich Supercomputing Centre and RWTH Aachen University
for providing computational resources.

\end{document}